\documentclass[fleqn,usenatbib]{mnras}

\usepackage{amsmath}
\usepackage{amssymb}

\usepackage{graphicx}
\usepackage{booktabs}
\usepackage[T1]{fontenc}
\usepackage{dblfloatfix}   

\graphicspath{{figures/}}

\newcommand{\vt}[1]{\mathbf{#1}}
\newcommand{\Lc}{L_{\mathrm c}}
\newcommand{\Nin}{N_{\mathrm{in}}}
\newcommand{\Mout}{M_{\mathrm{out}}}

\title[A discrete beam combiner for ELT petalling]{Sensing ELT petal modes with a discrete beam combiner: an end-to-end theoretical model}

\author[K. Madhav et al.]{%
K. Madhav,$^{1,2}$\thanks{E-mail: kmadhav@aip.de}
A.~S.~Nayak,$^{1}$
L.~Labadie$^{3}$
and R.~R.~Thomson$^{4}$
\\
$^{1}$Astrophotonics (innoFSPEC), Leibniz-Institut f\"ur Astrophysik Potsdam (AIP), An der Sternwarte 16, 14482 Potsdam, Germany\\
$^{2}$Institut f\"ur Physik und Astronomie, Universit\"at Potsdam, Karl-Liebknecht-Str. 24/25, 14476 Potsdam, Germany\\
$^{3}$I. Physikalisches Institut, Universit\"at zu K\"oln, Z\"ulpicher Str. 77, 50937 K\"oln, Germany\\
$^{4}$Institute of Photonics and Quantum Sciences, Heriot-Watt University, Edinburgh EH14 4AS, UK}

\date{Draft v9 -- \today}
\pubyear{2026}

\begin{document}
\label{firstpage}
\maketitle

\begin{abstract}
The European Southern Observatory's Extremely Large Telescope (ELT), the
$39$-m telescope under construction in Chile, forms its entrance pupil from
$798$ hexagonal segments that the six $0.5$-m-wide spider arms fragment
into six ``petals''. The low-wind / island effect, adaptive-optics
residuals, and mechanical drift imprint an independent piston on each
petal. These differential-piston (petal) modes remain largely undetected by
the ELT's baseline continuous-pupil wavefront sensors, a major bottleneck
for diffraction-limited, high-contrast imaging. This paper presents a
complete theoretical model of a single-mode photonic \emph{discrete beam
combiner} (DBC) configured to sense the six petal pistons. The model
constructs the coupled-mode description of a two-dimensional evanescently
coupled waveguide lattice with six inputs and forty-five outputs and
derives its visibility-to-pixel matrix (V2PM); treats the pupil remapping
that routes one sub-aperture per petal into the six inputs; develops the
five-dimensional petal-mode basis of the fragmented pupil; and propagates
the petal modes through the DBC to quantify the sensitivity. Optimising the
input-port spacing and interaction length by minimising the V2PM condition
number yields a well-conditioned device ($\mathrm{CN}=10.4$). It recovers
petal pistons with unit linear gain over the full $\pm\lambda/2$
($\pm 800$\,nm) unambiguous range and separates the five modes with
negligible cross-talk ($<10^{-3}$). For $1\%$ photometric noise the
reconstructed-piston error is $4.7$\,nm rms at $\lambda_{0}=1.6\,\mu$m.
Tested against petal amplitudes from end-to-end ELT simulations
($\approx40$--$300$\,nm rms atmospheric residual; $\sim0.3$--$1\,\mu$m
low-wind events), the DBC recovers each piston to a few nm rms at H band,
with a two-wavelength extension capturing the larger events. The DBC is
therefore a compact, chromatically robust petalometer for the fragmented
ELT pupil.
\end{abstract}


\section{Introduction}
The European Southern Observatory's Extremely Large Telescope (ELT), now
under construction on Cerro Armazones in the Atacama Desert of northern
Chile, has a primary mirror of diameter $D=39$\,m, with telescope first
light planned for $2029$
\citep{ESO2025}. Its entrance pupil is, however, not simply connected. The
primary mirror is a mosaic of $798$ hexagonal segments, and the six arms
of the structure supporting the $4.25$-m secondary mirror cast wide
shadows across the pupil. These six spider arms, each $\approx0.5$\,m wide,
divide the ELT pupil into six fragments of nearly equal area, the
\emph{petals} \citep{BertrouCantou2020}. The first-generation ELT
instruments MICADO, HARMONI and METIS all rely on single-conjugate
adaptive optics with a pyramid wavefront sensor and are therefore directly
exposed to the petalling problem addressed here
\citep{Hippler2019,Bond2022,BertrouCantou2022}.

Because the petals are optically isolated by the (opaque) spiders, a
piston applied to one petal is not constrained by its neighbours. Thermal
gradients in still air around the spiders (the \emph{low-wind effect},
seen already on the VLT/SPHERE, \citealt{Sauvage2016}), the so-called
\emph{island effect}, adaptive-optics (AO) reconstruction gaps over the
spiders, and slow thermo-mechanical drift all inject differential piston
(and, to a lesser extent, tip/tilt) between petals. These
\emph{petal modes} fragment the point spread function, destroy the
diffraction-limited core, and are a primary obstacle to high-contrast
imaging at an ELT \citep{BertrouCantou2022,Leboulleux2022}. Differential
piston is, moreover, poorly observed by continuous-pupil sensors. A
Shack--Hartmann sensor measures only local wavefront slopes, whereas a
differential piston is a constant phase offset with zero gradient
everywhere except at the spider gaps that the sensor cannot see; it
therefore produces almost no slope signal. A pyramid sensor, in turn,
suffers $2\pi$ confusion \citep{BertrouCantou2022}. A dedicated,
phase-unwrapped petal sensor is thus required. A range of such sensors has
been investigated for the ELT, from pyramid-based double-path and
spatially filtered schemes \citep{Levraud2022,Levraud2024} to focal-plane
approaches using co-phasing strategies \citep{Yang2025} or deep-learning
reconstruction \citep{JaninPotiron2025}.

Astrophotonics offers a natural route. A single-mode photonic
\emph{discrete beam combiner} (DBC) is a two-dimensional array of
evanescently coupled waveguides, written in a glass substrate by
ultrafast-laser inscription, that combines light from many inputs ``all
in one'' and encodes the mutual coherence of the inputs into a set of
output intensities \citep{Minardi2010,Minardi2012,Saviauk2013,Nayak2023}. The DBC is
intrinsically achromatic in design, occupies a few cubic millimetres, and
reads out on a small number of detector pixels. Laser-written DBCs have
been characterised in the laboratory across the near- and mid-infrared
\citep{Saviauk2013,Nayak2020Lband}, integrated monolithically with
three-dimensional pupil remappers
\citep{Nayak2020Hband,Piacentini2020}, and demonstrated on sky
\citep{Nayak2020WHT,Nayak2021AO,Nayak2022}. Originally proposed to
measure complex visibilities of multi-telescope interferometers
\citep{Minardi2010}, the same device, when fed with one sub-aperture per
petal of a single fragmented pupil, measures the mutual coherence between
petals (i.e.\ the differential pistons). This work extends earlier
multi-telescope, on-sky DBC demonstrations
\citep{Nayak2020WHT,Nayak2021AO,Nayak2022} by re-purposing the combiner as a dedicated
petalometer for the ELT's single fragmented aperture.

This paper builds the complete theory of a six-input DBC
petalometer, from coupled-mode propagation to petal sensitivity. The
paper is organised as four self-contained models, mirroring the physical
propagation chain:
Section~\ref{sec:dbc} develops the DBC coupled-mode model, transfer
matrix and V2PM, and establishes the minimum output count
$\Mout\ge\Nin^{2}$;
Section~\ref{sec:remap} models the pupil remapping from six petal
sub-apertures into the six DBC inputs;
Section~\ref{sec:petal} develops the petalling / differential-piston
model of the fragmented pupil; and
Section~\ref{sec:sens} propagates the six petal modes through the device
and evaluates the petal sensitivity. Results are summarised in
Section~\ref{sec:results} and concluded in Section~\ref{sec:concl}.

\section{Theoretical model of the discrete beam combiner}
\label{sec:dbc}

\begin{figure*}
\centering
\includegraphics[width=\textwidth]{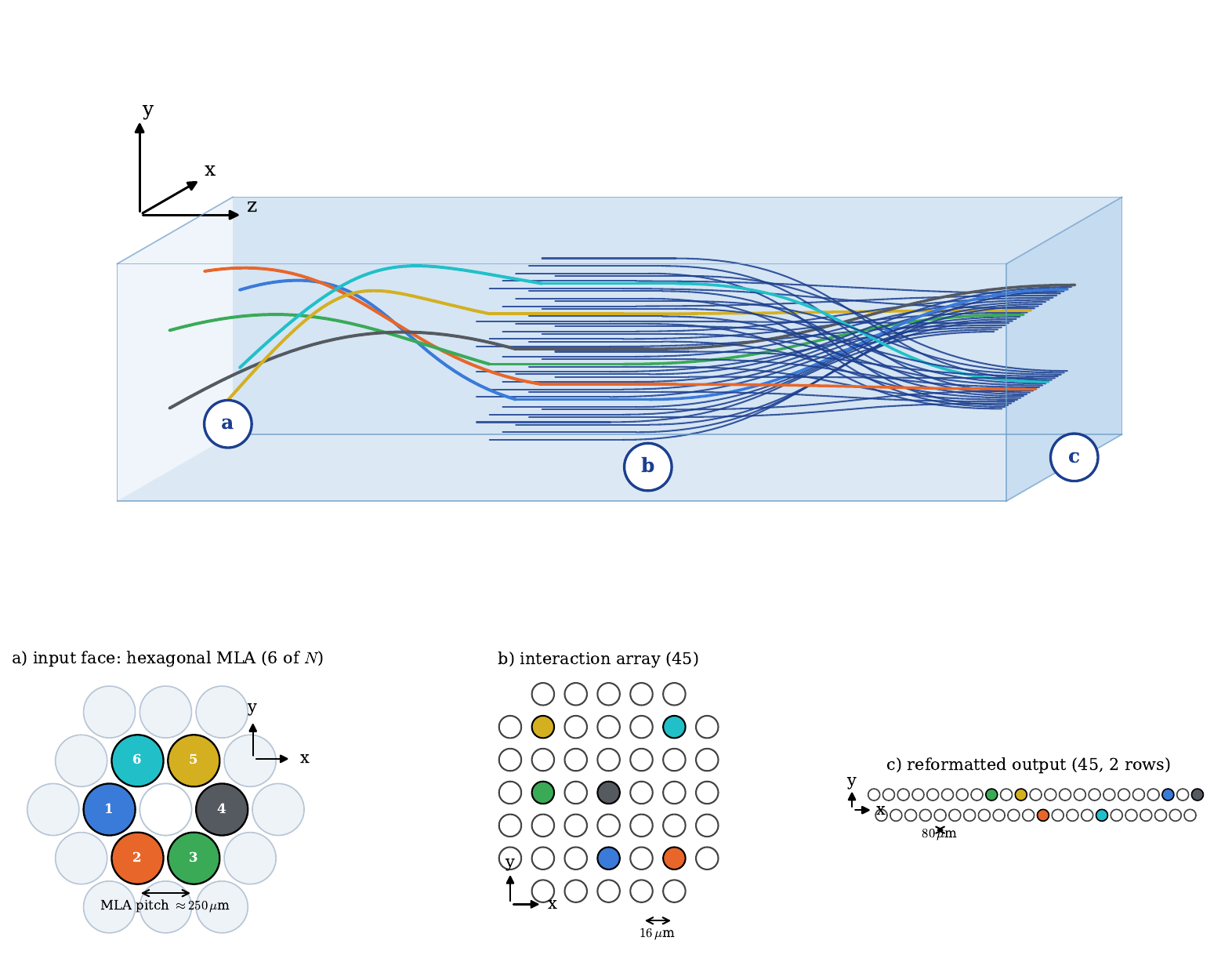}
\caption{Overview of the proposed monolithic six-input / forty-five-output
DBC petalometer, a single glass substrate written by ultrafast-laser
inscription. \emph{(a)} At the input face a hexagonal microlens array
(MLA) at a standard pitch ($\approx250\,\mu$m) couples one sub-aperture
per ELT petal into six single-mode waveguides (the six lenslets
neighbouring the obscured central one). \emph{(a$\to$b)} A
three-dimensional pupil remapper routes the six inputs, weaving in all
three axes inside the substrate, from the coarse MLA pitch down to the
dense interaction pitch ($\approx16\,\mu$m). \emph{(b)} The six inputs
join a two-dimensional array of $45$ evanescently coupled single-mode
waveguides (the DBC), where discrete diffraction mixes the
fields. \emph{(c)} The $45$ waveguides are reformatted into a two-row linear
output array at a pitch ($\approx80\,\mu$m) that can be imaged on a detector. Insets show the cross-sections at a), b) and c);
the six inputs are colour-coded consistently throughout.}
\label{fig:chip}
\end{figure*}

\subsection{Device architecture}
\label{sec:arch}
Figure~\ref{fig:chip} gives an overview of the device, which integrates
both pupil remapping and beam combination in a single laser-written glass
chip \citep{Saviauk2013,Nayak2020Hband,Piacentini2020,Nayak2023}. Light is
coupled in at the input face
(Fig.~\ref{fig:chip}a) by a hexagonal microlens array of standard pitch
($\approx250\,\mu$m): one lenslet samples one sub-aperture per petal, so the six color-coded segments of the MLA, reproduces the hexagonal
fragmentation of the ELT pupil (Section~\ref{sec:remap}). A three-dimensional single-mode pupil remapper
then carries the six beams through the substrate to a two-dimensional hexagonal lattice
(Fig.~\ref{fig:chip}b). This array features a $\approx16\,\mu$m pitch and comprises $\Mout=45$
evanescently coupled waveguides, where discrete diffraction redistributes the light and
encodes the inter-petal coherences (Sections~\ref{sec:dbc}--%
\ref{sec:sens}). Finally the $45$ waveguides are spatially reformatted into a compact two row linear array (Fig.~\ref{fig:chip}c) with an output pitch ($80\,\mu$m), convenient for
imaging onto a detector. The remainder of this section develops the
coupled-mode model of the interaction region and its retrieval matrix.

\subsection{Coupled-mode propagation in a 2D waveguide lattice}
Consider an array of $\Mout$ identical single-mode waveguides arranged on
a two-dimensional lattice and propagating along $z$. In the weak-guiding,
nearest-coupling regime the modal field amplitudes
$\vt{A}(z)=[A_1,\dots,A_{\Mout}]^{\mathrm T}$ obey the coupled-mode
equations \citep{Szameit2007,Nayak2023}
\begin{equation}
\mathrm{i}\,\frac{\mathrm d \vt{A}}{\mathrm d z}
   = \vt{C}\,\vt{A},
\label{eq:cme}
\end{equation}
where the (Hermitian) coupling matrix $\vt{C}$ has zero diagonal
(identical, phase-matched guides, $\Delta\beta=0$) and off-diagonal
elements set by the evanescent overlap of neighbouring modes,
\begin{equation}
C_{kl}=\frac{k_0^{2}}{2\beta}
   \frac{\displaystyle\int\!\!\int \psi_k^{*}\,\Delta n^{2}\,\psi_l\,
         \mathrm d x\,\mathrm d y}
        {\displaystyle\int\!\!\int \psi_k^{*}\psi_k\,
         \mathrm d x\,\mathrm d y},
\label{eq:kappa}
\end{equation}
with $\psi_k$ the fundamental mode of guide $k$, $\beta$ the propagation
constant and $\Delta n^{2}$ the index perturbation of the neighbouring
guide. On a square lattice with pitch $a$, two interactions are retained:
the nearest-neighbour (orthogonal, separation $a$) coupling $\kappa_1$ and
the next-nearest-neighbour (diagonal, separation $a\sqrt2$) coupling
$\kappa_2=r\,\kappa_1$, so that
\begin{equation}
C_{kl}=
\begin{cases}
\kappa_1, & |\vt r_k-\vt r_l| = a,\\[2pt]
r\,\kappa_1, & |\vt r_k-\vt r_l| = a\sqrt2,\\[2pt]
0, & \text{otherwise.}
\end{cases}
\end{equation}
The diagonal (next-nearest) coupling is essential: it lifts the
degeneracy of the discrete diffraction pattern and is what makes the
retrieval problem well conditioned \citep{Minardi2010,Nayak2023}.

For a pair of isolated guides Eq.~\eqref{eq:cme} gives the familiar
periodic power exchange with coupling length
$\Lc=\pi/(2\kappa_1)$, the length over which power transfers completely
from one guide to the other. The device length is therefore expressed in
units of $\Lc$.

Integrating Eq.~\eqref{eq:cme} gives an energy-conserving map
from the input amplitudes to the amplitudes at length $L$,
\begin{equation}
\vt{A}(L)=\vt{U}\,\vt{A}(0), \qquad
\vt{U}=\exp(-\mathrm{i}\,\vt{C}\,L),
\label{eq:U}
\end{equation}
where $\vt{U}$ is unitary because $\vt{C}$ is Hermitian. Light is
injected into $\Nin=6$ selected sites (the DBC \emph{inputs}); the full
$\Mout$ guides are read out. Writing $\vt{E}=[E_1,\dots,E_{\Nin}]^{\rm T}$
for the input input field ($E_i$), the output field at output m is the linear superposition
$N_{in}$ input fields given by: 

\begin{equation}
E_m=\sum_{i=1}^{\Nin} U_{mi}\,E_i ,
\label{eq:Em}
\end{equation}
where $U_{mi}$ matrix has a dimension of $\Mout\times\Nin$.

\subsection{Output intensities and the visibility-to-pixel matrix}
The measured quantity is the time-averaged power at each output,
$P_m=\langle|E_m|^{2}\rangle$. Expanding Eq.~\eqref{eq:Em}
\citep{Minardi2010,Nayak2023},
\begin{equation}
\begin{split}
P_m=\sum_{i=1}^{\Nin}|U_{mi}|^{2}\,\Gamma_{ii}
   +2\!\!\sum_{i<j}\!\big[&\Re(U_{mi}U_{mj}^{*})\,\Re\Gamma_{ij}\\
   &-\Im(U_{mi}U_{mj}^{*})\,\Im\Gamma_{ij}\big],
\end{split}
\label{eq:Pm}
\end{equation}
where $\Gamma_{ij}=\langle E_i E_j^{*}\rangle$ is the mutual coherence
(complex visibility) between inputs $i$ and $j$ and
$\Gamma_{ii}=\langle|E_i|^{2}\rangle$ the photometry. The phase of
$\Gamma_{ij}$ is exactly the differential piston between inputs $i$ and
$j$, which is the signal to be recovered.

\begin{figure}
\centering
\includegraphics[width=\columnwidth]{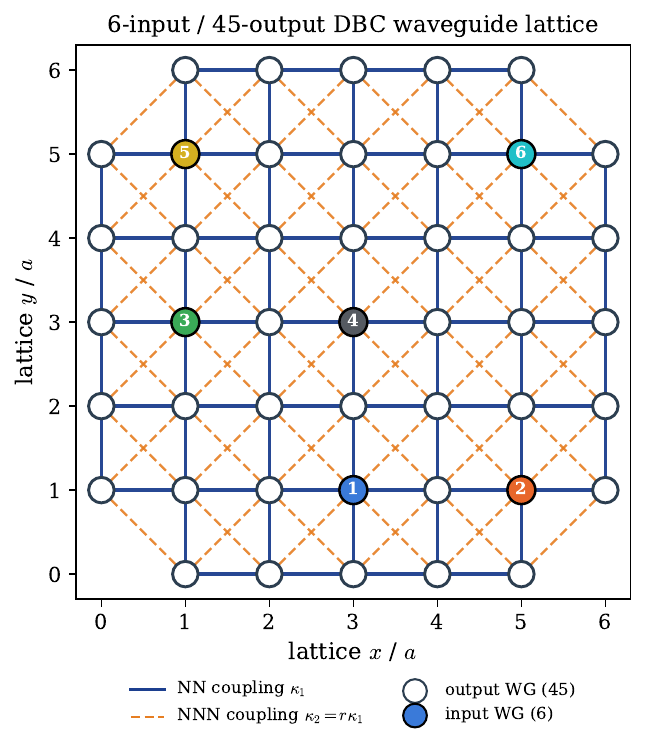}
\caption{Optimised six-input / forty-five-output DBC waveguide lattice.
Open circles are output waveguides; filled red circles (numbered 1--6)
are the injection sites. Heavy and light links denote the
nearest-neighbour ($\kappa_1$) and next-nearest-neighbour
($\kappa_2=r\kappa_1$) couplings retained in $\vt{C}$. The injection
sites and their spacing were optimised to minimise the V2PM condition
number. It was found that the injection sites achieved a minimum port spacing of $2a$, when the two coupling coefficients are
equal ($r=1$).}
\label{fig:lattice}
\end{figure}

Equation~\eqref{eq:Pm} is linear in the coherence terms, so it can be
written as the matrix relation
\begin{equation}
\vt{P}=\vt{V}\,\vt{J},
\label{eq:V2PM}
\end{equation}
where $\vt{P}$ is the $\Mout$-vector of output powers, $\vt{J}$ collects
the independent real coherence parameters,
\begin{multline}
\vt{J}=\big(\Gamma_{11},\dots,\Gamma_{\Nin\Nin},\,
   \Re\Gamma_{12},\dots,\Re\Gamma_{\Nin-1\,\Nin},\\
   \Im\Gamma_{12},\dots,\Im\Gamma_{\Nin-1\,\Nin}\big)^{\mathrm T},
\end{multline}
and $\vt{V}$ is the \emph{visibility-to-pixel matrix} (V2PM), equivalently
the transfer matrix of the DBC. The number of independent real parameters
is
\begin{equation}
N_{\rm coh}=\underbrace{\Nin}_{\Gamma_{ii}}
  +\,2\underbrace{\tfrac12\Nin(\Nin-1)}_{\Re\Gamma_{ij}\,,\,\Im\Gamma_{ij}}
  =\Nin^{2}.
\label{eq:Ncoh}
\end{equation}
\textbf{Minimum number of outputs.} The Equation~\eqref{eq:V2PM} is
invertible only if $\vt{V}$ has rank $N_{\rm coh}=\Nin^{2}$, which
requires at least as many independent output measurements,
\begin{equation}
\boxed{\;\Mout \ge \Nin^{2}\;}
\end{equation}
For $\Nin=6$ this gives a theoretical minimum of $\Mout=36$ output ports.
In practice the design over-samples to gain noise robustness; the adopted
value is $\Mout=45$ (a $7\times7$ square lattice with the four corner sites
removed, Fig.~\ref{fig:lattice}), giving $\Mout/N_{\rm coh}=1.25$.

\subsection{Coherence retrieval and conditioning}
\begin{figure}
\centering
\includegraphics[width=\columnwidth]{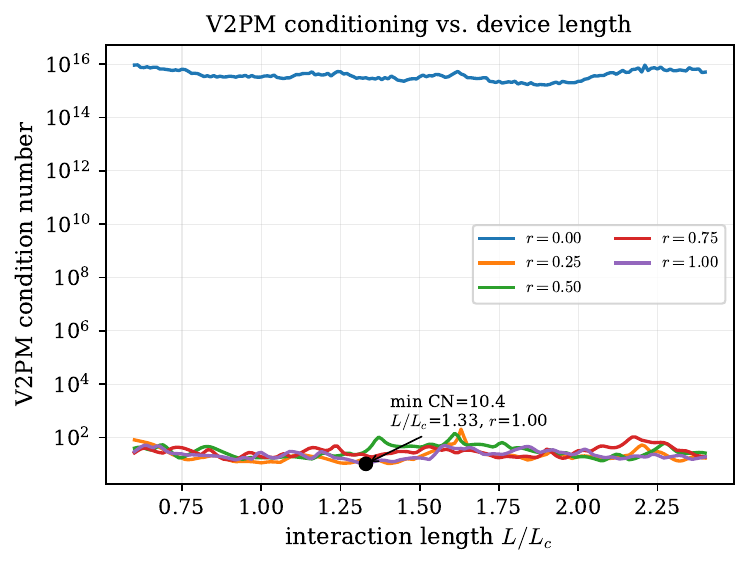}
\caption{V2PM condition number versus interaction length $L/\Lc$ for
several diagonal/orthogonal coupling ratios $r=\kappa_2/\kappa_1$, at the
optimised injection layout. The minimum CN${}=10.4$ is reached at
$L=1.33\,\Lc$ with equal couplings $r=1$. Without diagonal coupling
($r=0$) the matrix becomes singular at several lengths.}
\label{fig:cn}
\end{figure}

\begin{figure}
\centering
\includegraphics[width=\columnwidth]{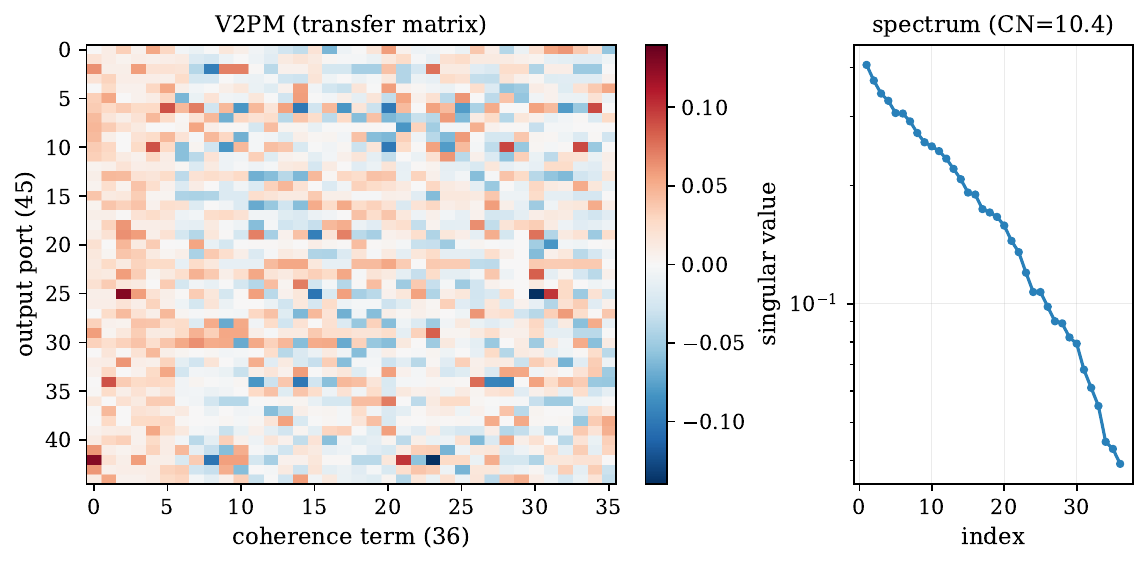}
\caption{Left: the optimised V2PM ($45\times36$), columns ordered as the
six photometric terms followed by the real and imaginary parts of the
fifteen mutual-coherence terms. Right: its singular-value spectrum; the
ratio of the extreme values is the condition number ($10.4$).}
\label{fig:v2pm}
\end{figure}

Because $\Mout>N_{\rm coh}$, Eq.~\eqref{eq:V2PM} is over-determined and is
inverted in the least-squares sense by the Moore--Penrose pseudo-inverse,
the pixel-to-visibility matrix $\vt{V}^{+}$ (P2VM),
\begin{equation}
\hat{\vt{J}}=\vt{V}^{+}\vt{P},\qquad
\vt{V}^{+}=(\vt{V}^{\mathrm T}\vt{V})^{-1}\vt{V}^{\mathrm T}.
\end{equation}
The robustness of the retrieval is governed by the condition number of
$\vt{V}$,
\begin{equation}
\mathrm{CN}(\vt{V})=\frac{\sigma_{\max}(\vt{V})}{\sigma_{\min}(\vt{V})},
\end{equation}
the ratio of the largest to smallest singular values. It bounds how much
the pseudo-inverse amplifies measurement errors: a relative perturbation
$\epsilon$ in the output $\vt{P}$ maps to a relative error of at most
$\mathrm{CN}\cdot\epsilon$ in the recovered coherence vector $\hat{\vt J}$,
because the components of $\vt{P}$ along the smallest singular direction
(which carry the weakest signal) are divided by the smallest singular
value during inversion and so are magnified the most.
The free design parameters are the interaction length
$L/\Lc$, the coupling ratio $r=\kappa_2/\kappa_1$, and, crucially,
the placement (spacing) of the six injection sites. Following the DBC
optimisation strategy of \citet{Minardi2010} and \citet{Nayak2023}, the
input-port layout is treated as a design variable, and the V2PM condition
number is jointly minimised over the configuration, $r$, and
$L/\Lc\in[0.6,2.4]$. Figure~\ref{fig:cn} shows the conditioning of the
optimised configuration versus length for several $r$. With no diagonal
coupling ($r=0$) the V2PM is rank-deficient (CN$\to\infty$) for many
lengths, confirming that next-nearest-neighbour coupling is mandatory.
The global optimum is $\mathrm{CN}=10.4$ at $L=1.33\,\Lc$ with
\emph{equal} couplings $r=1$ ($\kappa_2=\kappa_1$), recovering the
classic DBC result that the lowest condition number is reached when the
nearest and next-nearest coupling strengths are matched
\citep{Nayak2023}. The optimal injection sites, on lattice coordinates
$(3,1),(5,1),(1,3),(3,3),(1,5),(5,5)$, have a minimum port spacing of
$2a$ (Fig.~\ref{fig:lattice}). The optimised
transfer matrix is unitary to machine precision (residual
$\|\vt U^{\dagger}\vt U-\vt I\|_\infty\sim9\times10^{-16}$) and routes all
injected power to the output set. The optimised V2PM and its
singular-value spectrum, whose spread from largest to smallest value sets
the condition number and reflects how evenly the coherence terms are
encoded across the outputs, are shown in Fig.~\ref{fig:v2pm}.

\subsection{Polychromatic operation}
For a source of finite bandwidth $\Delta\lambda$ about $\lambda_0$, the
photometrically corrected fringe term at output $m$ acquires the
band-smearing envelope \citep{Nayak2023}
\begin{equation}
P_m \propto |U_{mi}U_{mj}^{*}|\;
   \frac{\sin\!\big(\pi\delta\,\Delta\lambda/\lambda_0^{2}\big)}
        {\pi\delta\,\Delta\lambda/\lambda_0^{2}}\;
   \cos\!\big(2\pi\delta/\lambda_0+\phi_{mij}\big),
\label{eq:poly}
\end{equation}
where $\delta=c\tau$ is the optical-path difference and $\phi_{mij}$ the
intrinsic phase of $U_{mi}U_{mj}^{*}$. Equation~\eqref{eq:poly} sets the
coherence envelope within which the (achromatic-by-design) V2PM
description holds; for the H band ($\lambda_0=1.6\,\mu$m,
$\Delta\lambda=0.15\,\mu$m) the envelope is wide compared with the
petal-piston excursions of interest ($\lesssim\lambda$), so the
narrow-band V2PM is an excellent approximation.

\section{Pupil remapping: six petals to six inputs}
\label{sec:remap}
The DBC requires its inputs to be fed by single spatial modes. The
fragmented telescope pupil (Fig.~\ref{fig:chip}a) is therefore re-imaged
and one sub-aperture per petal is injected into the six DBC waveguides
(Fig.~\ref{fig:chip}b), following the three-dimensional single-mode
pupil-remapper approach demonstrated for DBCs by \citet{Nayak2020Hband}
and \citet{Piacentini2020}. The ELT-like
pupil is modelled as a circle of diameter $D=39$\,m with a $28\%$ central obscuration,
divided into six $60^{\circ}$ petals by six $0.50$-m radial spiders
\citep{BertrouCantou2020} (Fig.~\ref{fig:remap}, left).

\subsection{Sub-aperture sampling and single-mode injection}
A small sub-aperture is selected within each petal and imaged onto the
core of the corresponding input waveguide. The coupling (injection)
efficiency is the overlap integral between the sub-aperture's
diffraction pattern and the fundamental waveguide mode. For a circular
sub-aperture matched to a Gaussian $\mathrm{LP}_{01}$ mode the ideal
overlap is $\rho_{\max}\simeq0.81$ \citep{Shaklan1988}; residual AO
aberration reduces this by the sub-aperture Strehl ratio $S$,
\begin{equation}
\rho_{\rm inj}\simeq\rho_{\max}\,S .
\end{equation}
For $S=0.5$ (a conservative ELT single-conjugate-AO value at the
sub-aperture scale) $\rho_{\rm inj}\simeq0.41$. The single-mode filtering
is precisely what converts seeing-induced amplitude/phase ripple within a
petal into a well-defined complex amplitude $E_i=a_i e^{\mathrm i\varphi_i}$
at input $i$ and phase $\varphi_i$ carries the petal piston.

\subsection{Uniform one-per-segment remapping}
\begin{figure}
\centering
\includegraphics[width=\columnwidth]{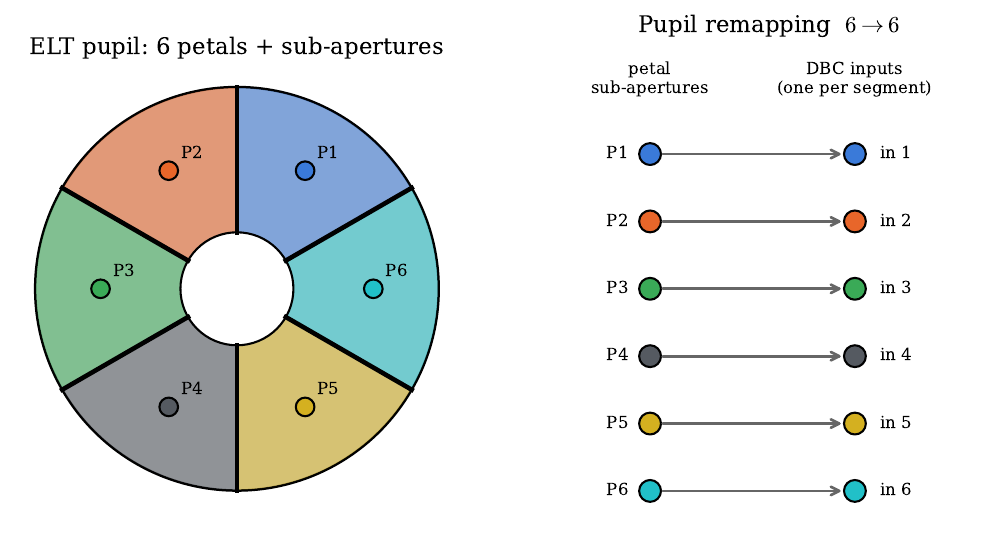}
\caption{Left: model ELT pupil fragmented into six petals by six spider
arms, with one sub-aperture placed uniformly at the centre of each
segment. Right: the $6\to6$ pupil remapping that injects each petal
sub-aperture into one dedicated DBC input waveguide; every segment is
sampled exactly once.}
\label{fig:remap}
\end{figure}

The fragmentation imposes a natural, symmetric sampling: \emph{one
sub-aperture per segment}, placed on the azimuthal mid-line of each
$60^{\circ}$ petal at the segment area-centroid radius
($R\simeq0.68\,R_{\rm pup}$; Fig.~\ref{fig:remap}). This
guarantees that every one of the six petals is sampled exactly once and
feeds one dedicated DBC input, so the petal-piston signal is measured
with maximal symmetry and no segment is left unmonitored.

The six sub-apertures define $\binom{6}{2}=15$ baselines, and a regular
hexagonal layout is geometrically redundant, containing only nine distinct
baseline vectors. In conventional aperture masking such redundancy is
damaging: pairs of sub-apertures with identical baselines overlap on the
detector, their fringe patterns add incoherently, and the differential
phase is lost. A DBC avoids this. Light from each petal enters its own
single-mode waveguide before being mixed in the photonic lattice, and the
visibility-to-pixel matrix (V2PM) recovers the phase difference of every
petal pair from the output intensities, unaffected by geometric
redundancy. The fifteen inter-petal coherences, and hence the differential
pistons, are thus recovered independently of pupil-baseline redundancy.
The uniform one-per-segment layout is therefore both the most symmetric
and a fully sufficient choice for petal sensing, and it is adopted
throughout this paper.

\subsection{Sub-aperture placement within a segment}
\label{sec:placement}
Pupil-remapping interferometers that \emph{image} an astronomical source,
including the laser-written DBC demonstrations of
\citet{Nayak2020Hband} and \citet{Nayak2020WHT}, deliberately choose
\emph{non-redundant} sub-aperture masks, because each baseline must sample
a distinct sky spatial frequency for aperture synthesis. Petalometry is a
different measurement: it senses the internal differential pistons of a
single fragmented wavefront, not the sky. Whether a non-redundant layout
helps here too is therefore worth asking. It does not, for two
reasons. First, the V2PM, and hence the condition number ($10.4$) and
the per-baseline coherence precision, is fixed by the waveguide array
and the input-site placement, and is \emph{independent of the pupil
geometry} (the sub-aperture coordinates never enter $\vt{U}$). Second, a
petal piston is uniform across its segment, so a single-mode sub-aperture
returns the same phase wherever it is placed: for ideal pistons the
location is immaterial and pupil-baseline redundancy is benign.

\begin{figure}
\centering
\includegraphics[width=\columnwidth]{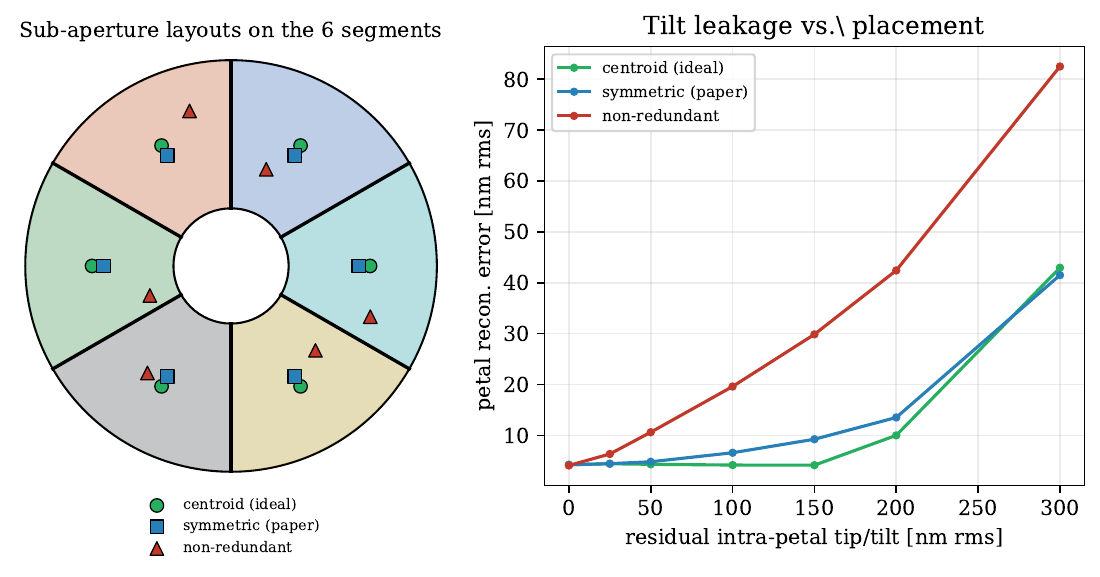}
\caption{Sub-aperture placement on the six ELT segments. \emph{Left:} the
three layouts compared: centroid, symmetric (adopted), and a
non-redundant mask. \emph{Right:} petal reconstruction error versus
residual intra-petal tip/tilt; centroid and symmetric placement stay near
the photometric-noise floor, whereas the non-redundant layout leaks tilt
into the measured piston and degrades steeply.}
\label{fig:subap}
\end{figure}

The placement does matter once \emph{residual intra-petal tip/tilt} is
present. A sub-aperture at position $\vt s_i$ reports the local mean phase
$p_i+\vt g_i\!\cdot\!\vt s_i$, whereas the physical petal piston is the
segment mean $p_i+\vt g_i\!\cdot\!\vt c_i$; the sensing error is
$\vt g_i\!\cdot\!(\vt s_i-\vt c_i)$. Here $p_i$ is the petal piston (an
optical-path difference, in nm), $\vt g_i$ the intra-petal wavefront
gradient (the full first-order \emph{tip/tilt} of segment $i$, a
two-component vector $(g_x,g_y)$ in nm per unit pupil radius), and
$\vt s_i$, $\vt c_i$ the two-dimensional positions (in pupil-radius units)
of the sub-aperture centre and the segment area-centroid, so that
$\vt g_i\!\cdot\!(\vt s_i-\vt c_i)$ is an optical-path error in nm. Because a
single sub-aperture per petal supplies exactly enough degrees of freedom to
fix the six pistons (five differential modes) and none to sense the
tip/tilt, this term is an irreducible bias: it vanishes at the centroid and
grows as the sub-aperture is moved off-centroid, which is exactly what a
non-redundant layout requires.
This is quantified with the optimised DBC using the same forward model and
P2VM inversion as Section~\ref{sec:sens}, the sole addition being the
injection model above. In each Monte-Carlo trial, independent petal
pistons $p_i\sim\mathcal{N}(0,\sigma_p^2)$ ($\sigma_p=130$\,nm rms) and
isotropic intra-petal tilts $\vt g_i\sim\mathcal{N}(0,\sigma_t^2)$
(\emph{both} orthogonal tip and tilt components) are drawn; the injected
fields $E_i=a_i\,e^{\mathrm{i}\,2\pi(p_i+\vt g_i\cdot\vt s_i)/\lambda_0}$
are formed and propagated through the device (Eq.~\ref{eq:forward}), $1\%$
photometric noise is added, the petal modes are reconstructed, and the rms
of the recovered pistons against the physical segment-mean values
$p_i+\vt g_i\!\cdot\!\vt c_i$ is computed
(Table~\ref{tab:placement}, Fig.~\ref{fig:subap}). For a pure piston
($\sigma_t=0$) the wavefront is flat across each segment, so the sampled
phase is independent of $\vt s_i$ and all layouts coincide at the
$\approx4$\,nm photometric-noise floor: redundancy is benign. Under a
realistic $100$\,nm rms intra-petal tilt, however, the non-redundant layout
degrades to $\sim20$\,nm rms while centroid/symmetric placement stays at
$4$--$7$\,nm, because a non-redundant mask must place the sub-apertures at
distinct radii and off-bisector azimuths to render the fifteen baseline
vectors distinct, displacing them from the centroids (mean offset $0.21$
versus $0.06\,R_{\rm pup}$) and amplifying the tilt leakage by
$\sim3$--$4\times$. That advantage, $15$ versus $9$ distinct baseline
\emph{vectors}, is a $u$--$v$-imaging metric with no bearing on
petalometry, since each petal feeds its own single-mode input and the DBC
measures every $\Gamma_{ij}$ on dedicated V2PM channels irrespective of the
pupil-baseline geometry. Each sub-aperture is therefore placed at its
segment centroid: redundancy is benign, and non-redundancy is not merely
unnecessary but mildly detrimental.

\begin{table}
\centering
\caption{Petal reconstruction error (rms, at $1\%$ photometry, $130$\,nm
rms pistons) for three sub-aperture layouts on the six segments, versus
residual intra-petal tip/tilt. The condition number ($10.4$) is identical
for all layouts.}
\label{tab:placement}
\small
\setlength{\tabcolsep}{4pt}
\begin{tabular}{lcccc}
\toprule
Layout & Distinct & Pure & $+100$\,nm & $+200$\,nm\\
       & baselines & piston & tilt & tilt\\
\midrule
Centroid (ideal)   & 9  & $4.2$\,nm & $4.4$\,nm  & $10.0$\,nm\\
Symmetric (adopted)& 9  & $4.2$\,nm & $6.6$\,nm  & $13.5$\,nm\\
Non-redundant      & 15 & $4.1$\,nm & $19.6$\,nm & $42.4$\,nm\\
\bottomrule
\end{tabular}
\end{table}

\section{Petalling and differential piston in fragmented pupils}
\label{sec:petal}
Attention now turns from the instrument to the petal-mode wavefront it
must sense. Let petal $i$ carry a uniform piston
phase $\varphi_i$ (in radians at $\lambda_0$), related to an optical-path
piston $p_i$ by $\varphi_i=2\pi p_i/\lambda_0$. After single-mode
filtering the field injected into input $i$ is
\begin{equation}
E_i=a_i\,e^{\mathrm{i}\varphi_i},\qquad i=1,\dots,6 ,
\label{eq:e_field}
\end{equation}
with $a_i$ the (real) photometric amplitude.

\subsection{The petal-mode basis}
A piston common to all six petals is an unobservable global phase; only
\emph{differential} pistons are physical. Modelling each segment as a single scalar piston, the petal state is the
vector $\boldsymbol{\varphi}=(\varphi_1,\dots,\varphi_6)^{\mathrm T}\in\mathbb{R}^{6}$.
A common offset $\varphi_i\to\varphi_i+\mathrm{const}$ leaves every measured
coherence $\Gamma_{ij}$ (whose phase is $\varphi_i-\varphi_j$) unchanged, so
the global-piston direction is exactly $\vt 1=(1,\dots,1)^{\mathrm T}$ and only
the subspace orthogonal to it is observable. Taking the six segments to be
equally weighted and using the Euclidean metric on $\mathbb{R}^{6}$, the
observable petal modes therefore span the mean-removed subspace, obtained by
the orthogonal projector. Defining the centering operator
\begin{equation}
\vt{H}=\vt{I}_6-\tfrac{1}{6}\vt{1}\vt{1}^{\mathrm T},
\end{equation}
its eigenvectors with non-zero eigenvalue form an orthonormal basis
$\{\vt b_1,\dots,\vt b_5\}$ of the five differential-piston modes; the
sixth (null) eigenvector is the global piston. For segments of unequal flux or area the uniform mean $\tfrac{1}{6}\vt 1\vt 1^{\mathrm T}$
is replaced by a weighted projector $\vt 1\vt w^{\mathrm T}$ with $\sum_i w_i=1$. Any petal state decomposes
as
\begin{equation}
\boldsymbol{\varphi}=\bar\varphi\,\vt 1+\sum_{k=1}^{5}c_k\,\vt b_k ,
\end{equation}
where $\bar\varphi=\tfrac{1}{6}\sum_i\varphi_i$ is the global piston, the
$\{\vt b_k\}_{k=1}^{5}$ are the orthonormal eigenvectors of $\vt H$ with unit
eigenvalue, and $c_k=\vt b_k^{\mathrm T}\boldsymbol{\varphi}$. Since
$\{\vt 1/\sqrt6,\vt b_1,\dots,\vt b_5\}$ is a complete orthonormal basis of
$\mathbb{R}^{6}$, this decomposition is exact and unique. It rests on the same
assumptions as Eq.~(16): each segment carries a single scalar piston
(pure-piston model), the piston state is quasi-static over one exposure, and
the six segments are equally weighted.

The DBC constrains only the five coefficients $c_k$, for two related reasons.
First, the device is read out in intensity, $P_m=\langle|E_m|^2\rangle$, which
is invariant under a common phase shift $\boldsymbol{\varphi}\to
\boldsymbol{\varphi}+\mathrm{const}\,\vt 1$; the global piston $\bar\varphi$ is
therefore unobservable in principle. Second, all phase information is carried
by the mutual coherences $\Gamma_{ij}$, whose phases are the pairwise
differences
\begin{equation}
\arg\Gamma_{ij}=\varphi_i-\varphi_j
=\sum_{k=1}^{5}c_k\,(b_{k,i}-b_{k,j}),
\end{equation}
in which the $\bar\varphi\,\vt 1$ term cancels identically. These differences
span exactly the five-dimensional mean-removed subspace, so the fifteen
measured baseline phases over-determine the five $c_k$ while leaving
$\bar\varphi$ entirely unconstrained; it lies in the null space of the
measurement. This
five-dimensional petal space is the target of the sensor.

\subsection{Origin and amplitude of the petal modes}
\label{sec:amp}
Physically the $c_k$ are driven by three mechanisms whose amplitudes are
now well characterised by end-to-end ELT simulations.
\emph{(i) Atmospheric residual petal.} Because the thick spiders hide the
wavefront locally, the SCAO loop cannot fully enforce phase continuity
across them, leaving a residual differential piston. Simulations of the
ELT pupil (six $0.50$-m spiders, $28\%$ central obstruction, $798$
segments) give a residual petal of $\approx40$\,nm rms at good seeing
($r_0\approx15$\,cm), rising to $\approx300$\,nm rms in poor seeing
($r_0\approx10$\,cm) \citep{BertrouCantou2020,Levraud2024}.
\emph{(ii) Low-wind / island effect.} Radiative cooling of the spiders in
still air imprints slowly varying petal steps; observed and modelled
amplitudes reach $\sim300$\,nm and up to $\sim1\,\mu$m OPD around the
spiders, substantially larger than the atmospheric residual
\citep{Milli2018,Sauvage2016,Levraud2024}.
\emph{(iii) Reconstruction null space and drift.} Petal modes lie close
to the poorly sensed subspace of continuous-pupil reconstructors, and
thermo-mechanical drift adds a slow component
\citep{Bonnefond2016,Schwartz2018}.
Temporally, most of the residual-petal power lies below $\approx10$\,Hz,
so a dedicated petal loop running at $\gtrsim50$--$100$\,Hz is sufficient
\citep{BertrouCantou2020}. A petal sensor must therefore be
\emph{sensitive} at the nm level, \emph{fast} ($\sim100$\,Hz), and
\emph{unambiguous} over the few-hundred-nm to $\mu$m range spanned by
these regimes.

\section{Propagation of petal modes and petal sensitivity}
\label{sec:sens}

\begin{figure*}[t]
\centering
\includegraphics[width=\textwidth]{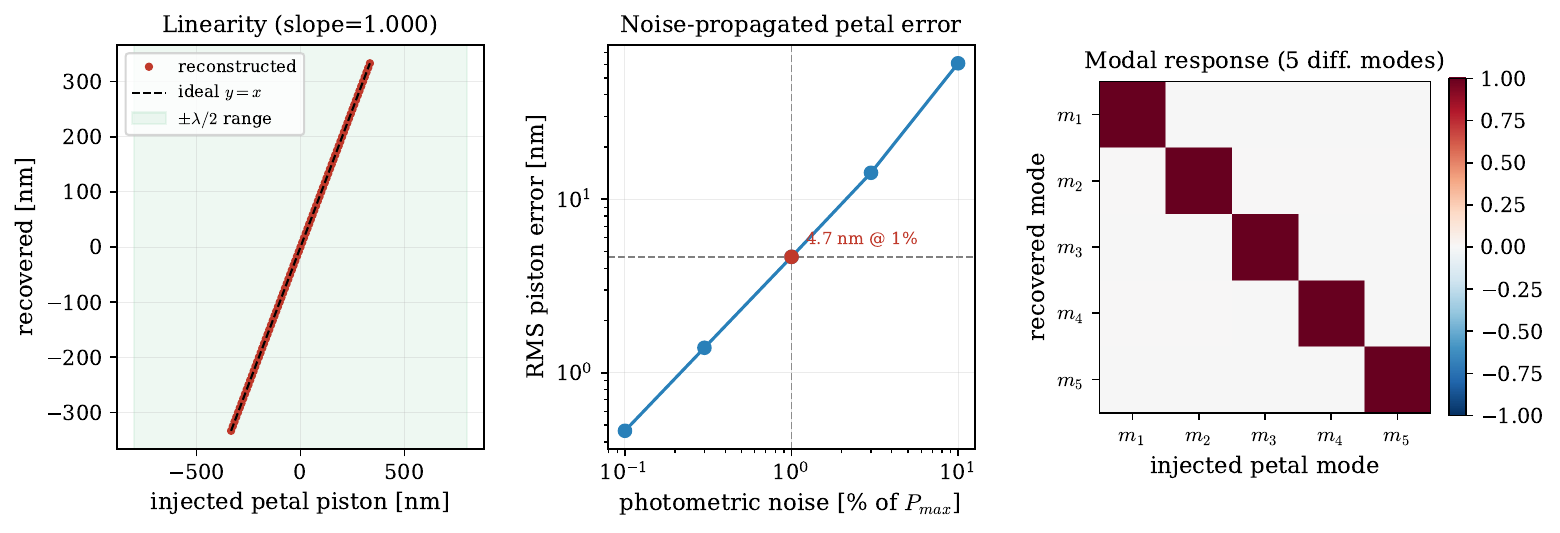}
\caption{Petal sensitivity of the optimised six-input DBC. Left:
reconstructed versus injected petal piston, showing unit gain over the
$\pm\lambda/2$ unambiguous range. Centre: rms reconstructed-piston error
versus photometric noise, scaling linearly ($4.7$\,nm at $1\%$). Right:
the $5\times5$ modal-response matrix for the five differential-piston
modes, an identity to within noise.}
\label{fig:sens}
\end{figure*}

\subsection{Forward model}
The petal field propagating through the DBC follows from substituting
Eq.~\ref{eq:e_field} into Eq.~\ref{eq:Em}; squaring the result gives the
output-power vector as a function of the six petal pistons,
\begin{equation}
P_m(\boldsymbol{\varphi})=
  \Big|\sum_{i=1}^{6}U_{mi}\,a_i\,e^{\mathrm i\varphi_i}\Big|^{2}.
\label{eq:forward}
\end{equation}
With equal photometry ($a_i=1$) the total detected power equals the total
injected power, $\sum_{m}P_m=\sum_i|E_i|^2=6$ (one unit per input), since the
lossless array propagator $\vt U$ is unitary and its six injected columns are
orthonormal; summed over all $45$ outputs this confirms energy conservation.
Here $a_i$ denotes the amplitude already coupled into waveguide $i$, the
front-end injection efficiency $\rho_{\rm inj}$ (Section~\ref{sec:remap})
being excluded.

\subsection{Linear petal sensitivity}
The local petal sensitivity is the Jacobian
$J_{mk}=\partial P_m/\partial\varphi_k$ evaluated at the operating point.
Projected onto the differential-mode basis, $\tilde{\vt J}=\vt J\,\vt B\in\mathbb{R}^{45\times5}$,
where $\vt J\in\mathbb{R}^{45\times6}$ is the Jacobian of Eq.~(18) and
$\vt B=[\vt b_1\,\dots\,\vt b_5]\in\mathbb{R}^{6\times5}$ has the five orthonormal
differential-piston modes as its columns, mapping the five petal modes to the
$45$ output-power sensitivities. Its singular values
$\{0.58,0.50,0.40,0.35,0.32\}$ are all comparable, with a
\emph{response} condition number of $1.82$: every differential mode
produces a similar, well-measured photometric signal, so no petal mode is
weakly sensed. The piston is encoded into the fringe (cosine) terms of
Eq.~\eqref{eq:Pm}, giving maximum slope at quadrature, as in a
conventional ABCD estimator but realised simultaneously over all fifteen
baselines.

\subsection{Reconstruction and linearity}
Petal pistons are recovered in three steps: (a) apply the P2VM to the
output powers to obtain $\hat{\vt J}$ and hence each
$\Gamma_{ij}$; (b) read the differential piston of each baseline from
the visibility phase, $\arg(\Gamma_{ij})=\varphi_i-\varphi_j$; and (c)
solve the over-determined linear system relating the fifteen baseline
phases to the five differential-piston modes by least squares. Because
the phase is obtained directly (not from a fitted fringe amplitude), the
estimator is linear: injecting a piston on one petal and reconstructing it
gives a gain $=1$ over the full unambiguous range
$|\,p\,|<\lambda/2=800$\,nm (Fig.~\ref{fig:sens}, left). Beyond
$\pm\lambda/2$ the phase wraps, setting the single-wavelength capture
range; multi-wavelength operation (Eq.~\ref{eq:poly}) extends it via the
group delay.

\subsection{Noise propagation}
Gaussian photometric noise of standard deviation $\sigma=\eta\,P_{\max}$
is added to every output and the petal state reconstructed, for random
petal states of $50$\,nm rms. Here
$P_{\max}$ is the peak (brightest) output-port power and $\eta=\sigma/P_{\max}$
is the dimensionless fractional noise level (e.g.\ $\eta=0.01$ for $1\%$
photometry), representative of a detector-referred noise near full scale. The reconstructed-piston error grows linearly
with $\eta$ (Fig.~\ref{fig:sens}, centre): from $0.48$\,nm at
$\eta=0.1\%$ to $4.7$\,nm at $\eta=1\%$ and $60$\,nm at $\eta=10\%$. The $1\%$ result, $\sigma_p\simeq4.7$\,nm rms at $\lambda_0=1.6\,\mu$m,
follows from the linear error-propagation chain of Section~\ref{sec:dbc}:
a fractional output-power error $\eta$ is amplified by the pseudo-inverse of
the V2PM, by at most its condition number ($\mathrm{CN}=10.4$), into an
error on the retrieved coherences $\Gamma_{ij}$ and hence on their phases
$\arg\Gamma_{ij}=\varphi_i-\varphi_j$; the phase-to-piston factor
$\lambda_0/2\pi$ (a phase error $\delta\varphi$ corresponds to an optical-path
error $\delta p=\lambda_0\,\delta\varphi/2\pi$) then converts this into a
piston. Schematically $\sigma_p\sim(\lambda_0/2\pi)\,g(\mathrm{CN})\,\eta$, so
the reconstruction error scales linearly with $\eta$ and is minimised by the
low condition number of the optimised device. A precision of $\eta=1\%$ is
representative of a $16$-bit detector operating near full well, where read and
quantisation noise together amount to $\sim1\%$ of the peak port signal
$P_{\max}$.

\subsection{Modal cross-talk}
A final check verifies that the five petal modes are independently
measured. Each orthonormal differential mode is injected in turn,
reconstructed, and projected back onto the basis. The resulting $5\times5$
modal-response matrix is an identity to within noise (off-diagonal rms
$\sim10^{-15}$, diagonal mean $1$; Fig.~\ref{fig:sens}, right), confirming
that the device introduces no internal cross-talk between petal modes and
that residual errors are set only by photometric noise.

\begin{figure*}[t]
\centering
\includegraphics[width=\textwidth]{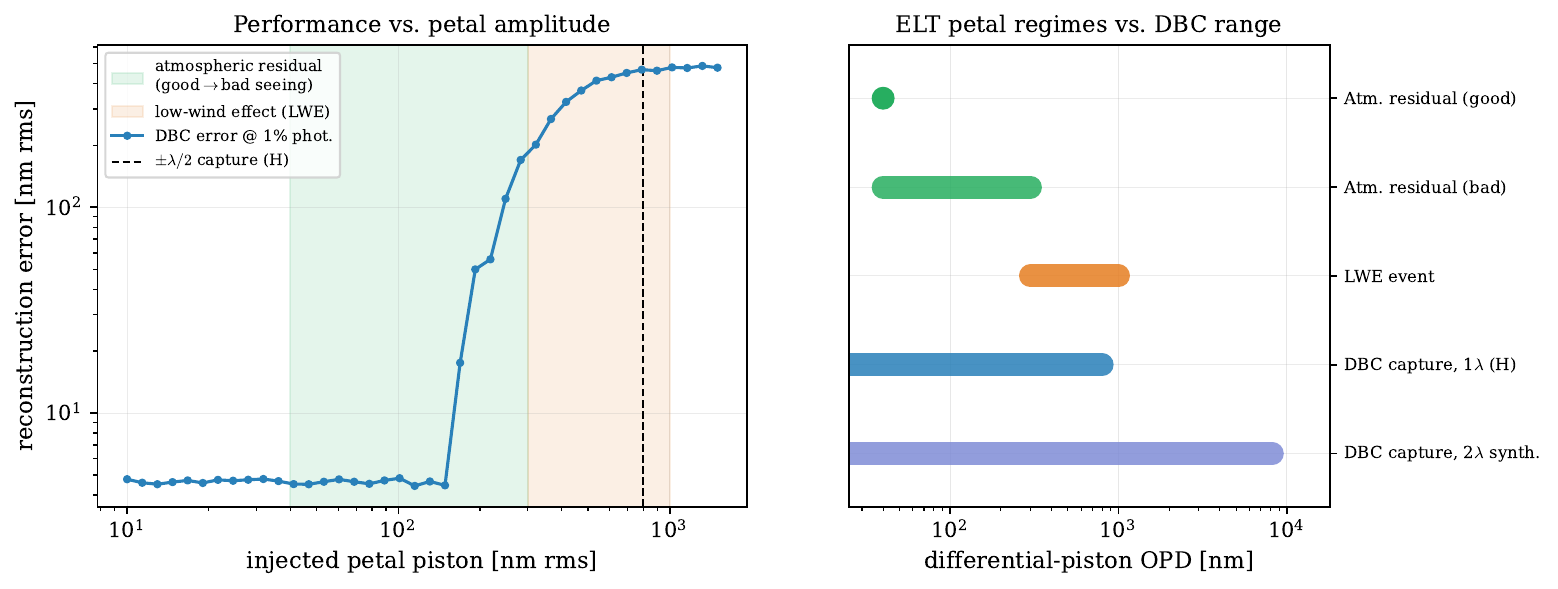}
\caption{DBC petal-sensing performance against the literature-grounded ELT
petal regimes. \emph{Left:} reconstructed-piston error (at $1\%$
photometric noise) versus the rms amplitude of injected random petal
states. The error stays at the $\approx4.5$\,nm floor across the
atmospheric-residual band (green; good$\to$bad seeing,
$40$--$300$\,nm rms \citealt{BertrouCantou2020,Levraud2024}) until the differential excursions begin to exceed the single-wavelength
$\pm\lambda/2$ capture (dashed); beyond this the phase wraps and the
reconstruction no longer tracks the true piston, so the error rises steeply
and then \emph{saturates at a plateau} (of order $\lambda/2$) set by the
spread of the wrapped phase, independent of the injected amplitude.
\emph{Right:} the literature petal regimes (atmospheric residual and
low-wind effect, LWE) compared with the DBC unambiguous range at H band
($\pm\lambda/2=\pm800$\,nm) and with a two-wavelength synthetic range that
covers the full LWE regime.}
\label{fig:regimes}
\end{figure*}

\subsection{Performance across the ELT petal regimes}
\label{sec:regimes}
A final test compares the DBC against the realistic petal amplitudes
catalogued in Section~\ref{sec:amp}. For random petal states of increasing
rms amplitude reconstructed at $1\%$ photometric noise
(Fig.~\ref{fig:regimes}), the error remains at the
$\approx4.5$\,nm floor throughout the atmospheric-residual regime, with
$4.6$\,nm rms at good seeing ($40$\,nm) and $4.5$\,nm at a representative
mid-seeing residual ($130$\,nm), confirming that for routine SCAO
operation the DBC is photometric-noise limited and meets the nm-level
requirement. As the per-petal rms grows the \emph{differential} excursions
between petals approach the single-wavelength capture range
($\pm\lambda/2=800$\,nm at H), and the estimate degrades through phase
wrapping; at bad seeing ($300$\,nm rms) the single-wavelength error rises
to $\approx190$\,nm. This is not a limitation of the combiner but of the
monochromatic phase ambiguity: dispersing the DBC outputs and combining
two or more wavelengths synthesises a group-delay observable whose
unambiguous range extends to several $\mu$m
(Eq.~\ref{eq:poly}; \citealt{Saviauk2013}), comfortably covering the LWE
regime ($300$\,nm--$1\,\mu$m). The photonic readout is effectively
instantaneous, so a petal loop at the required $\sim100$\,Hz
(Section~\ref{sec:amp}) is trivially supported. Animations of the $45$
output-waveguide intensities and of the corresponding ELT petal-mirror
movements, for all five regimes of Fig.~\ref{fig:regimes}, are provided as
online supporting information.

\begin{figure*}[t]
\centering
\includegraphics[width=\textwidth]{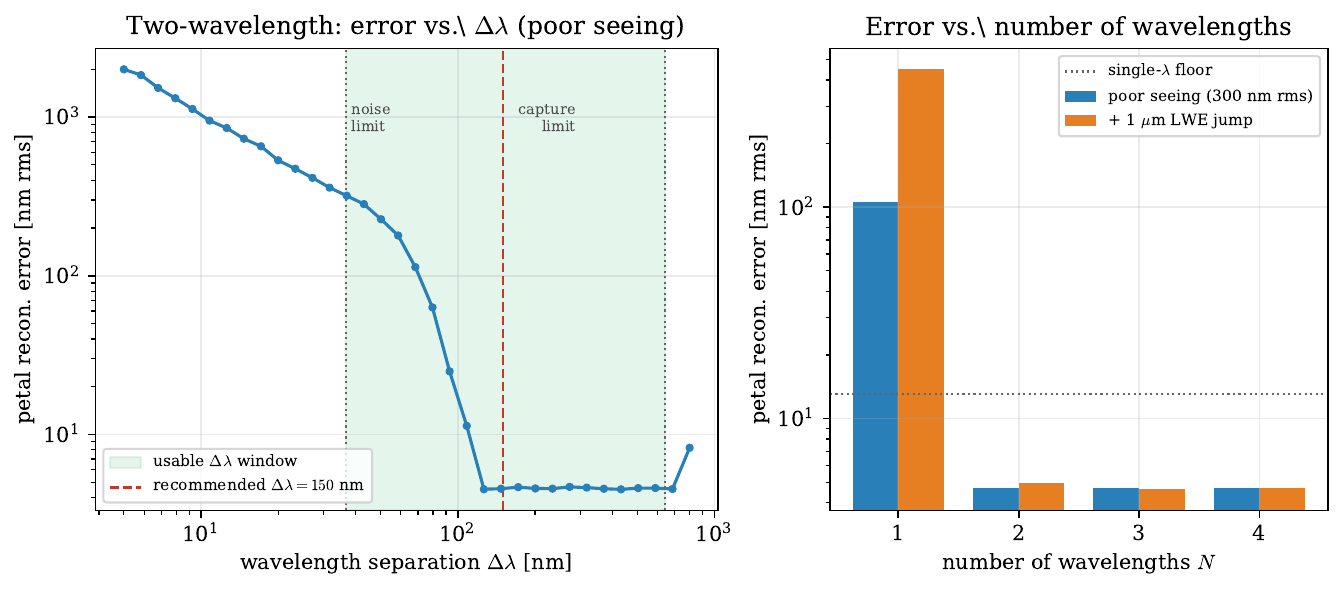}
\caption{Two-wavelength capture-range analysis for ELT poor seeing.
\emph{Left:} petal reconstruction error versus wavelength separation
$\Delta\lambda$ for a pair centred at $1.6\,\mu$m, under poor-seeing petal
states ($300$\,nm rms). The usable window (green) is bounded below by phase
noise ($\Delta\lambda\gtrsim40$\,nm, where the synthetic-OPD error reaches
$\lambda/2$) and above by the synthetic capture range
($\Delta\lambda\lesssim640$\,nm); within it the error returns to the
single-wavelength floor. The recommended $\Delta\lambda=150$\,nm (H-band
edges) is marked. \emph{Right:} petal error versus number of wavelengths
$N$, for poor seeing and for poor seeing plus a $1\,\mu$m LWE jump; a single
pair ($N=2$) already collapses the $N=1$ wrapping errors ($106$ and
$449$\,nm) to the $\approx4.7$\,nm floor (dotted).}
\label{fig:twowl}
\end{figure*}

\subsection{Two-wavelength capture for the poor-seeing and LWE regimes}
\label{sec:twocolour}
The single-wavelength wrapping seen at poor seeing
(Fig.~\ref{fig:regimes}) is removed by sensing at two or more wavelengths
and forming the synthetic wavelength
$\Lambda=\lambda_1\lambda_2/|\lambda_2-\lambda_1|$, whose unambiguous range
$\pm\Lambda/2$ greatly exceeds $\pm\lambda/2$
\citep{Cheng1985,Saviauk2013}. A coarse estimate from $\Lambda$ resolves
the fringe order of the fine single-$\lambda$ measurement, so the capture
range is set by $\Lambda$ while the precision stays at the
single-wavelength floor. Two requirements bound the separation
$\Delta\lambda$:
\begin{equation}
\underbrace{\Delta\lambda \le \frac{\lambda_0^{2}}{2\,\mathrm{OPD}_{\max}}}
   _{\text{capture: }\Lambda\ge2\,\mathrm{OPD}_{\max}}
\qquad\text{and}\qquad
\underbrace{\Delta\lambda > 2\sqrt2\,\sigma_{\rm OPD}}
   _{\text{noise: }\sigma_\Lambda<\lambda/2},
\label{eq:dlam}
\end{equation}
where $\sigma_\Lambda=\sqrt2\,(\Lambda/\lambda_0)\,\sigma_{\rm OPD}$ is the
propagated synthetic-OPD error. For the ELT poor-seeing case
($\mathrm{OPD}_{\max}\approx2\,\mu$m, covering the bad-seeing $3\sigma$
differential plus a $1\,\mu$m LWE event) and the per-baseline precision
$\sigma_{\rm OPD}\approx13$\,nm, Eq.~\eqref{eq:dlam} gives a usable window
$\Delta\lambda\approx40$--$640$\,nm at $\lambda_0=1.6\,\mu$m; robust
operation (keeping $\sigma_\Lambda\lesssim\lambda/5$) favours
$\Delta\lambda\gtrsim100$\,nm.

The adopted pair is the H-band edges $\lambda_1=1.50$, $\lambda_2=1.65\,\mu$m
($\Delta\lambda=150$\,nm), giving $\Lambda=16.5\,\mu$m, an unambiguous
range of $\pm8.2\,\mu$m, and $\sigma_\Lambda\approx190$\,nm
($\ll\lambda/2$). Figure~\ref{fig:twowl} confirms the behaviour: outside
the window the estimate fails (noise blow-up at small $\Delta\lambda$,
capture loss at large $\Delta\lambda$), while within it the poor-seeing
error returns to the $\approx4.7$\,nm floor. A \emph{single pair} ($N=2$)
already restores full precision in both the poor-seeing and the
poor-seeing-plus-LWE cases, collapsing the single-wavelength errors of
$106$ and $449$\,nm to $\approx4.7$\,nm. A third wavelength adds an
intermediate synthetic rung for extra margin (fainter guide stars, or two
channels confined to one band) but is not required at H band. In summary,
\emph{two wavelengths separated by $\gtrsim100$\,nm (recommended
$\Delta\lambda\simeq150$\,nm) suffice} to cover the full ELT petal range at
poor seeing, with three recommended for operational margin; since the DBC
outputs can be spectrally dispersed, these channels come at no extra
component cost.

\section{Results and discussion}
\label{sec:results}

\begin{table}
\centering
\caption{Device and reconstruction parameters of the six-input DBC
petalometer ($\lambda_0=1.6\,\mu$m, H band).}
\label{tab:summary}
\small
\setlength{\tabcolsep}{4pt}
\begin{tabular}{lc}
\toprule
Device / reconstruction & Value\\
\midrule
Input ports $\Nin$ (= ELT petals)      & 6\\
Independent coherence terms $\Nin^{2}$  & 36\\
Minimum output ports                    & 36\\
Adopted output ports $\Mout$            & 45\\
Lattice                                 & $7\times7$ $-$ 4 corners\\
Min.\ input-port spacing (optimised)    & $2a$\\
Optimal coupling ratio $\kappa_2/\kappa_1$ & 1.0\\
Optimal interaction length $L/\Lc$      & 1.33\\
V2PM condition number                   & 10.4\\
Unitarity residual                      & $9\times10^{-16}$\\
Pupil sampling                          & 1 per segment\\
Distinct baseline vectors               & 9 of 15\\
Ideal single-mode injection $\rho_{\max}$ & 0.81\\
Differential-piston (petal) modes       & 5\\
Petal-response condition number         & 1.82\\
Linearity gain                          & 1.000\\
Capture range (H, single $\lambda$)     & $\pm800$\,nm\\
Modal cross-talk (off-diag.\ rms)       & $<10^{-3}$\\
\bottomrule
\end{tabular}
\end{table}

\begin{table}
\centering
\caption{Literature ELT petal regimes and the corresponding DBC
performance.}
\label{tab:regsummary}
\small
\setlength{\tabcolsep}{4pt}
\begin{tabular}{lc}
\toprule
Petal regimes \& DBC performance & Value\\
\midrule
Atm.\ residual petal (good$\to$bad)$^{a}$ & $40\to300$\,nm rms\\
Low-wind-effect amplitude$^{b}$         & $0.3$--$1\,\mu$m\\
Recon.\ error, good seeing ($40$\,nm)   & $4.6$\,nm rms\\
Recon.\ error, mid seeing ($130$\,nm)   & $4.5$\,nm rms\\
Recon.\ error, bad seeing$^{c}$         & $\approx190$\,nm rms\\
Required petal-loop rate$^{a}$          & $\sim50$--$100$\,Hz\\
Two-$\lambda$ pair (recommended)        & $1.50$ \& $1.65\,\mu$m\\
Synthetic wavelength $\Lambda$          & $16.5\,\mu$m ($\pm8.2\,\mu$m)\\
Usable $\Delta\lambda$ window           & $40$--$640$\,nm\\
Recon.\ error, bad seeing, $N{=}2$      & $4.7$\,nm rms\\
\bottomrule
\end{tabular}
\\[3pt]
{\footnotesize $^{a}$\citet{BertrouCantou2020};
$^{b}$\citet{Milli2018,Levraud2024};
$^{c}$single-wavelength wrapping, removed by two-wavelength synthesis.}
\end{table}

Tables~\ref{tab:summary} and~\ref{tab:regsummary} collect the key model
parameters and results.
The six-input DBC meets every requirement of an ELT petalometer. It uses
the theoretical minimum information channel ($\Nin^{2}=36$ coherence
terms) over-sampled by a compact $45$-output lattice; after optimising the
input-port spacing it is well conditioned (CN${}=10.4$), comparable to
the best four-input devices in the literature \citep{Nayak2023}; it is
linear with unit gain and
unambiguous over $\pm800$\,nm; it reaches few-nm reconstruction ($4.7$\,nm rms at 1\% photometry); and it separates the five physical petal modes
without cross-talk. Tested against the petal amplitudes from end-to-end
ELT simulations it remains photometric-noise limited (a few nm) across the
atmospheric-residual regime, with the larger low-wind events accessible
through a two-wavelength extension of the capture range
(Section~\ref{sec:regimes}).

Unlike continuous-pupil sensors, the DBC measures all fifteen inter-petal
baselines simultaneously and extracts the differential piston directly
from the visibility phase, bypassing the $2\pi$ phase ambiguity that limits
pyramid wavefront sensors \citep{BertrouCantou2022}. Its principal
costs are the photon budget of sub-aperture sampling
($\rho_{\rm inj}\sim0.4$) and the need for a one-time laboratory
calibration of the V2PM. The narrow-band approximation
(Eq.~\ref{eq:poly}) and the assumption of pure petal piston (neglecting
intra-petal tip/tilt) are the main idealisations; both are relaxable by
spectral dispersion of the outputs and by sampling more than one
sub-aperture per petal, and will be addressed in future work together
with an end-to-end AO-residual simulation.

\section{Conclusions}
\label{sec:concl}
This paper has presented a complete, end-to-end theoretical model of a
single-mode discrete beam combiner configured to sense petalling and
differential piston in the fragmented pupil of the ELT. The framework
integrates four modelling steps: the coupled-mode DBC physics
($\Mout\ge\Nin^{2}$), the $6\to6$ pupil remapping, the five-mode piston
basis, and the beam propagation through the chip. A compact six-input,
$45$-output DBC linearly recovers all five ELT petal modes with
nanometre-level precision at percent-level photometry and zero modal
cross-talk. Tested against literature petal amplitudes, the
device is photometric-noise limited ($\approx4.7$\,nm) across the
atmospheric-residual regime at H band; for the poor-seeing and low-wind
regimes, where single-wavelength wrapping appears, a two-wavelength
synthetic-wavelength scheme (two channels separated by
$\gtrsim100$\,nm, recommended $1.50$ \& $1.65\,\mu$m, with three for
margin) restores the few-nm precision over the full
$\sim\mu$m petal range. The DBC is therefore a strong candidate
petalometer for the ELT. Building on the on-sky DBC demonstrations of
\citet{Nayak2021AO} and \citet{Nayak2022}, a dedicated laboratory
demonstration of this framework, with a closed-loop simulation including
AO residuals, is the natural next step.

\section*{Acknowledgements}
This work draws on the Astrophotonics (innoFSPEC) programme at AIP and BMFTR-funded campaigns (grants 03Z22A511, 03Z22AN11, 03Z22AB1A, 03Z22AI1), and partial support was provided by the PICS4SENS project, funded by the State of Brandenburg through the Investitionsbank des Landes Brandenburg (ILB), with support from the European Regional Development Fund (ERDF/EFRE), grant number 86000879.

\section*{Data availability}
The simulation code that generates all results, figures and animations in
this paper, together with the animations themselves, are available from
the corresponding author on reasonable request.

\section*{Supporting Information}
The following animations are available as online supporting information,
covering the five ELT petal regimes of Fig.~\ref{fig:regimes}:
\begin{description}
\item[\texttt{regimes\_all.mp4}] the $45$ output-waveguide intensities on
  the interaction-region lattice (Fig.~\ref{fig:chip}b), for the five
  regimes evolving in synchrony;
\item[\texttt{pupil\_output\_all.mp4}] the ELT petal-mirror movements (six
  segments shaded by piston) shown alongside the $45$ intensities in the
  two-row reformatted readout (Fig.~\ref{fig:chip}c);
\end{description}
with individual per-regime versions of each also provided. The six petal
pistons are driven by a smooth, band-limited temporal signal representative
of the $\lesssim10$\,Hz petal dynamics, scaled to each regime's amplitude.

\bibliographystyle{mnras}
\bibliography{references}

\bsp
\label{lastpage}
\end{document}